# Spurious symmetry-broken phase in a bidirectional two-lane ASEP with narrow entrances：A perspective from mean field analysis and current minimization principle

Bo Tian[1], Rui Jiang[2,*], Mao-Bin Hu[1], Bin Jia[2]

[1]*School of Engineering Science, University of Science and Technology of China, Hefei 230026, P. R. China*

[2]*School of Traffic and Transportation, Beijing Jiaotong University, Beijing 100044, P. R. China*

## Abstract

As one of the paradigmatic models of non-equilibrium systems, the asymmetric simple exclusion process (ASEP) has been widely used to study many physical, chemical, and biological systems. The ASEP shows a range of nontrivial macroscopic phenomena, among which, the spontaneous symmetry breaking has gained much attention. Nevertheless, as a basic problem, it has been controversial that whether there exist one or two symmetry-broken phases in the ASEP. Based on mean field analysis and current minimization principle, this paper demonstrates that one of the broken-symmetry phases does not exist in a bidirectional two-lane ASEP with narrow entrances. Moreover, an exponential decay feature is observed, which has been used to predict phase boundary in the thermodynamic limit. Our findings might be generalized to other ASEP models and thus deepen the understanding of the spontaneous symmetry breaking in non-equilibrium systems.

## I. Introduction

Statistical investigation of non-equilibrium systems is a tough task. To understand the basic statistical properties of non-equilibrium systems, many paradigmatic models have been proposed. As one of the paradigmatic models of non-equilibrium systems, the asymmetric simple exclusion process (ASEP) has been widely used to study many physical, chemical, and biological systems [1-5], such as biopolymerization [6], gel electrophoresis [7], the kinetics of synthesis of proteins [8-10], biological transport [11-16], polymer dynamics in dense media [17], diffusion through membrane channels [18], surface growth [19,20], glassy dynamics [21,22], traffic flow [23,24], and so on.

The basic ASEP model is defined on a one-dimensional discrete lattice with *L* sites that are either occupied by a single particle or empty. Particles move along the lattice obeying a hard-core exclusion principle. This simple model can reproduce many non-equilibrium phenomena such as spontaneous symmetry breaking [25-37], boundary-induced [38-40] and bulk-induced [41] phase transitions, phase separation and condensation [42-46], shock formation [11,12,41,47], and so on. Thus, as Blythe and Evans [4] pointed out, "*our interest in the ASEP lies in its having acquired the status of a fundamental model of nonequilibrium statistical physics in its own right in much the same way that the Ising model has become a paradigm for equilibrium critical phenomena.*"

The spontaneous symmetry breaking has been studied across various fields. The first ASEP model exhibiting spontaneous symmetry breaking is known as the "bridge model" [25,26]. From then on, the phenomenon in ASEP models has gained much attention [27-37]. However, mechanism of this phenomenon is still not well understood. In particular, there are controversial reports about the number of symmetry-broken phases in the bridge model. The existence of one of the symmetry-broken phases (i.e., asymmetric low density (LD)/low-density phase) has been disputed on the basis of computer simulation by Arndt et al. [37]. Extensive high-precision Monte Carlo simulations indicated that the phase may disappear in the thermodynamic limit [27].

---

[*] Electronic address: jiangrui@bjtu.edu.cn

In the bridge model, particles moving in opposite directions interact with each other at every site of the lattice. Thus, Pronina and Kolomeisky [28] argued that it is not clear how the interactions between these particles localized in the specific parts of the system will affect the symmetry breaking. Motivated by this fact, they have proposed to study the bidirectional two-lane asymmetric exclusion processes with narrow entrances, in which interactions of particles in the two channels only happen at the entrances [28]. The system has been analyzed using a mean-field theoretical approach and extensive Monte Carlo computer simulations at different system sizes. It has been shown via mean field analysis that the model can also reproduce two phases with broken symmetry, i.e., the asymmetric high-density (HD)/LD phase, and the asymmetric LD/LD phase.

To study the size-scaling dependence of the asymmetric LD/LD phase in their model, Pronina and Kolomeisky [28] carried out computer simulations for the systems with different sizes up to L = 12,000. It was shown that "*the region of existence for the asymmetric LD/LD phase seems to shrink constantly with increasing L without reaching a saturation, which suggests that the phase probably does not exist in the thermodynamic limit*". However, Pronina and Kolomeisky [28] also claimed that the numerical results are not very conclusive, thus more careful investigations of this phase are needed in order to understand symmetry breaking phenomenon.

In this paper, we study the issue whether the asymmetric LD/LD phase exists or not in the bidirectional two-lane ASEP with narrow entrances. We point out that in previous mean field analysis, the necessary conditions have been wrongly regarded as the sufficient conditions. Consequently, the asymmetric LD/LD phase was shown to exist. We demonstrate that the asymmetric LD/LD phase should not exist based on mean field analysis and current minimization principle.

The paper is organized as follows. The model is briefly reviewed in Section II. The mean field analysis is presented in Section III. Section IV summarizes the paper.

## II. Model

The sketch of the model is shown in Fig.1. The system consists of two parallel one-dimensional L lattices with particles moving in different lanes in the opposite directions. Each lattice site can be either empty, or occupied by one particle. Hopping between the lanes is not allowed. In the bulk, particles in lane 1 (2) hop to the right (left) with rate 1 if the next site is empty. At the entrance site, a particle is injected with rate $\alpha$, provided the site is empty and exit site at the other lane is unoccupied. At the exit site, a particle is removed with rate $\beta$.

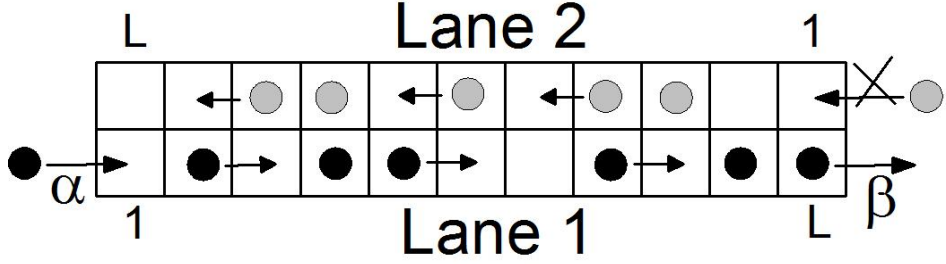

FIG. 1. Sketch of the model. Dark particles move along lane 1 from left to right, and gray particles move on lane 2 from right to left. Allowed transitions are shown by arrows. Crossed arrows indicate forbidden transitions.

### III. Mean field analysis

**A. Simple mean field**

Pronina and Kolomeisky [28] have presented a simple mean field analysis of the bidirectional two-lane system. They have shown that the conditions for existence of the asymmetric LD/LD phase are

$$\frac{\alpha}{1+\alpha+\alpha^2} < \beta < \frac{\alpha(1-2\alpha)+2\alpha\sqrt{\alpha^2-\alpha+1}}{3} \tag{1}$$

However, the conditions are actually the necessary conditions.

To determine whether the asymmetric LD/LD phase can exist or not, we resort to the current minimization principle proposed in [29,30,41]. Specifically, for the parameter set that yields more than one solution, the solution that has the minimum system current is the most stable one, and thus exists. Other solutions do not exist.

When the necessary condition of the existence of the asymmetric LD/LD phase is satisfied, the system current, which is the sum of the currents on the two lanes and denoted as $J_A$, can be calculated

$$J_A = \left(\frac{\beta}{\alpha}\right)^2 + 2\beta - \frac{\beta}{\alpha} \tag{2}$$

If the symmetric LD phase exists, then the system current, denoted as $J_B$, is

$$J_B = \frac{2\alpha^2\beta + \beta\sqrt{(\alpha+\beta)^2 - 4\alpha^2\beta} - \beta^2 - \alpha\beta}{\alpha^2} \tag{3}$$

Finally, if the system is in asymmetric HD/LD phase, then the system current, denoted as $J_C$, is

$$J_C = \beta(1-\beta) + \alpha\beta(1-\alpha\beta) \tag{4}$$

Fig.2 shows an example of the three currents at $\alpha = 0.4$. One can see that the curve of $J_A$ intersects the curve of $J_B$ at $\beta_R$ and the curve of $J_C$ at $\beta_L$. It can be easily proved that $\beta_L = \frac{\alpha}{1+\alpha+\alpha^2}$ and $\beta_R = \frac{\alpha(1-2\alpha)+2\alpha\sqrt{\alpha^2-\alpha+1}}{3}$. Thus, $J_A > J_C$ when $\beta$ is in the range of Eq.(1). This means that even if the necessary condition of the existence of the asymmetric LD/LD phase is satisfied, the phase does not exist due to current minimization principle. For general value of $\alpha$, the result does not change, which can be easily proved.

On the other hand, the curve of $J_B$ intersects the curve of $J_C$ at $\beta_c$. This means that when $\beta < \beta_c$, the system is in HD/LD phase. When $\beta > \beta_c$, the system is in symmetric LD phase.

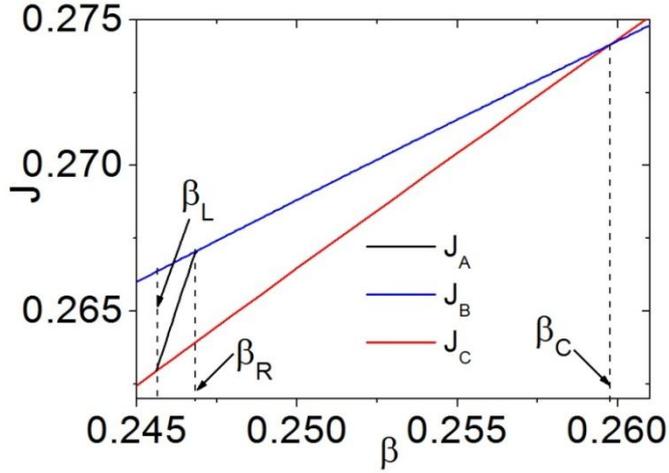

FIG. 2.The plot of $J_A$, $J_B$, and $J_C$ versus $\beta$, obtained from simple mean field analysis. The parameter $\alpha = 0.4$.

**B. Cluster mean field**

While simple mean field analysis ignores correlations between any two sites, the correlations in the cluster have been considered in cluster mean field analysis. We number the sites from the entrance site to the exit site as site 1,2,3,…,L, see Fig.1. In the N-cluster mean field analysis, we consider N consecutive sites (from exit site) on one lane and N consecutive sites (from entrance site) on the other lane. Fig.3 shows the example with $N = 1$. In this case, there are four possible states of the system. We define $P_{11}$ as a probability that the two sites are occupied, $P_{00}$ as a probability that the two sites are empty, $P_{10}$, $P_{01}$ as the probabilities that one site is empty and the other site is empty, respectively.

Due to conservation of probabilities, one has

$$P_{00} + P_{01} + P_{10} + P_{11} = 1 \tag{5}$$

Now we can write the master equations of the evolution of the four probabilities.

$$\frac{dP_{00}}{dt} = -(\alpha + p_1)P_{00} + \beta P_{01} + p_2 P_{10}$$

$$\frac{dP_{01}}{dt} = -\beta P_{01} + p_1 P_{00} + p_2 P_{11}$$

$$\frac{dP_{10}}{dt} = -(p_1 + p_2)P_{10} + \alpha P_{00} + \beta P_{11}$$

$$\frac{dP_{11}}{dt} = -(p_2 + \beta)P_{11} + p_1 P_{10}$$

In the stationary state, we have $\frac{dP_{00}}{dt} = \frac{dP_{01}}{dt} = \frac{dP_{10}}{dt} = \frac{dP_{11}}{dt} = 0$. Thus, one has

$$-(\alpha + p_1)P_{00} + \beta P_{01} + p_2 P_{10} = 0 \tag{6}$$
$$-\beta P_{01} + p_1 P_{00} + p_2 P_{11} = 0 \tag{7}$$
$$-(p_1 + p_2)P_{10} + \alpha P_{00} + \beta P_{11} = 0 \tag{8}$$
$$-(p_2 + \beta)P_{11} + p_1 P_{10} = 0$$

However, note that only three of the four equations are independent ones.

Next we can analyze the asymmetric LD/LD phase. We express the bulk density of lane 2 as

$$\rho_2 = P_{10} + P_{11} \quad (9)$$

We introduce $p_1$ as the hopping probability of a particle to the exit site of lane 1, and $p_2$ as the hopping probability of a particle at the entrance site of lane 2 (see Fig. 3). We assume

$$p_2 = 1 - \rho_2 \quad (10)$$

Note that this is the only assumption in the cluster mean field analysis.

Thus, in the asymmetric LD/LD phase, we have seven variables $P_{00}$, $P_{01}$, $P_{10}$, $P_{11}$, $p_1$, $p_2$, $\rho_2$ and six equations (5)-(10). In order to solve the equations, $p_1$ needs to be given firstly. We cannot obtain analytical result of the equations. Instead, we can obtain only numerical result. When the equations are solved, we can calculate the bulk density of lane 1 from $J_1 = \rho_1(1-\rho_1)$, and $J_1 = \rho_L \beta$, where $\rho_L = P_{11} + P_{10}$ is density on the exit site of lane 1. We would like to mention that the two sites at the other end of the system satisfy similar equations.

Note that in Ref.[35], we introduce the bulk density of lane 1 as $\rho_1$. Thus, the flow rate on lane 1 is $J_1 = \rho_1(1-\rho_1)$, so that $\rho_L = \rho_1(1-\rho_1)/\beta$. Based on $p_1(1-\rho_L) = J_1$, we calculate $p_1 = \frac{\beta \rho_1(1-\rho_1)}{\beta - \rho_1(1-\rho_1)}$. However, similar analytical relationship between $p_1$ and $\rho_1$ cannot be obtained in N-cluster case (N>1) due to the correlations in the cluster.

Next we consider the asymmetric HD/LD phase. Suppose lane 1 is in HD, one has $p_1 = 1 - \beta$. At the same time, Eqs.(5)-(10) are still satisfied. Thus, we have six variables $P_{00}$, $P_{01}$, $P_{10}$, $P_{11}$, $p_2$, $\rho_2$ and six equations. Solving the equations, one can obtain the system current.

N-cluster mean field analysis with N>1 can be carried out similarly and details will not be present here. With the increase of N, the only assumption (Eq.(10)) in the cluster mean field analysis becomes more and more accurate, but the number of system states and accordingly the number of master equations increase as $2^{2N}$.

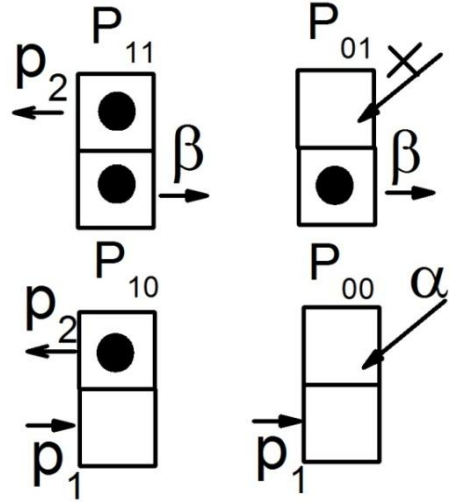

FIG. 3. Four possible states of the vertical cluster composed of the exit site of lane 1 and entrance site of lane 2 in one cluster mean field analysis.

*B.1 One-cluster mean field analysis*

Now we show the results of one-cluster mean field analysis. The black and red lines in Fig.4 (top panels) show examples of the curve of $\rho_1 = f_1(\rho_2)$ and $\rho_1 = f_2(\rho_2)$, obtained from the two sites at two ends of the system. Here $\rho_1$ and $\rho_2$ are bulk densities on lanes 1 and 2, respectively. To illustrate difference of the two curves more clearly, we show difference of the two curves $f_1(\rho_2) - f_2(\rho_2)$ in the bottom panels. It can be seen that the two curves intersect at three points A, B and C. Points B and C are symmetric about $\rho_1 = \rho_2$. Point A corresponds to $\rho_1 = \rho_2$. If one fixes $\alpha$ and decreases $\beta$, points B and C shift toward the two ends (Fig.4(b)). If one increases $\beta$, points B and C shift toward point A until becoming in superposition with point A (Fig.4(c)).

Thus, the necessary condition for the existence of asymmetric LD/LD phase is that there should exist the two points B and C, and $\rho_1 < \beta$ at point B and $\rho_2 < \beta$ at point C. Note that in Ref.[35], the necessary condition has been incorrectly presented as $\rho_1 < 1/2$ at point B and $\rho_2 < 1/2$ at point C.

We have compared the three currents $J_A$, $J_B$, and $J_C$. The results are the same as in simple mean field analysis: the asymmetric LD/LD phase does not exist.

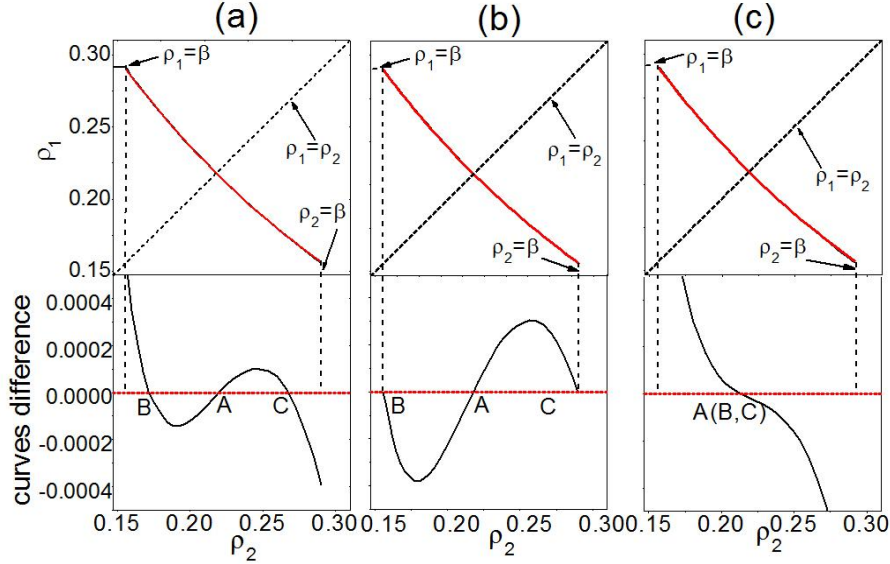

FIG. 4. The two curves of $\rho_1$ versus $\rho_2$ (top) and difference of the two curves (bottom). The parameter $\alpha = 0.8$, (a) $\beta = 0.3095$, (b) $\beta = 0.308436257$, (c) $\beta = 0.3106$.

*B.2. N-cluster mean field*

Next we perform N-cluster mean field analysis and increase N from 1. The result does not change until N = 4. However, when N > 4, the results become different. Fig.5 shows examples of the curves of $\rho_1$ vs. $\rho_2$ and $f_1(\rho_2) - f_2(\rho_2)$ in the 5-cluster mean field analysis. One can see that the two curves intersect at five points A-E. Points B and C are still symmetric about $\rho_1 = \rho_2$. Point A still corresponds to $\rho_1 = \rho_2$. Points D and E are also symmetric about $\rho_1 = \rho_2$. This means that there are three solutions: Points B and C correspond to one asymmetric LD/LD solution I; Points D and E correspond to the other asymmetric LD/LD solution II; Point A still corresponds to symmetric LD solution. If one fixes $\alpha$ and decreases $\beta$, points B and C shift toward two ends until coinciding with D and E, respectively, see Fig.5(b). With the further decrease of β, the asymmetric LD/LD solution vanishes, see Fig.5(c). If one increases β, points B and C shift toward point A until becoming in superposition with point A, see Fig.5(d). With the further increase of β, the asymmetric LD/LD solution vanishes too, see Fig.5(e).

We compare the flow rates of asymmetric HD/LD solution, asymmetric LD/LD solution I, asymmetric LD/LD solution II and symmetric LD solution, see Fig.6. In the range of asymmetric LD/LD solution $\beta_L < \beta < \beta_R$, in which the necessary condition of the existence of asymmetric LD/LD phase is satisfied, the flow rates of two asymmetric LD/LD solutions are both larger than that of asymmetric HD/LD solution. Changing the value of α, the result does not change, either.

We have increased N until N=6, the result does not change anymore. Although we are not able to prove rigorously, the mean field analysis and the current minimization principle strongly demonstrate that the asymmetric LD/LD phase does not exist.

Fig.7 shows the boundaries obtained from simple mean field, cluster mean field analysis, and simulations. Note that since the asymmetric LD/LD phase still exists in the simulations, we show the boundary separating asymmetric HD/LD phase from asymmetric LD/LD phase. The boundary separating asymmetric LD/LD phase from symmetric LD phase is not shown in the simulation results. One can see that cluster mean field analysis performs better than simple mean field analysis. With the increase of N, the analytical boundaries approach the simulation ones.

Finally we study the dependence of phase boundary on N in N-cluster mean field analysis. Fig.8(a) shows that at a given α, the boundary separating asymmetric HD/LD phase from symmetric LD phase is remarkably consistent with the exponential decay $\beta_c - \beta_{c,\infty} \propto \text{Exp}(a\text{N} + b)$. Therefore, one expects that $\beta_c = \beta_{c,\infty}$ when N → ∞. Fig.8(b) and (c) show that the boundary separating symmetric HD phase from the other two phases also exhibit exponential decay feature. The predicted boundaries in the limit N→ ∞ based on exponential decay feature are also shown in Fig.7. Simulations in larger system size need to be carried out to validate the prediction. However, one needs to be very careful with the simulations in very large systems due to that the random-number generator produces pseudorandom number series instead of true random number series [48].

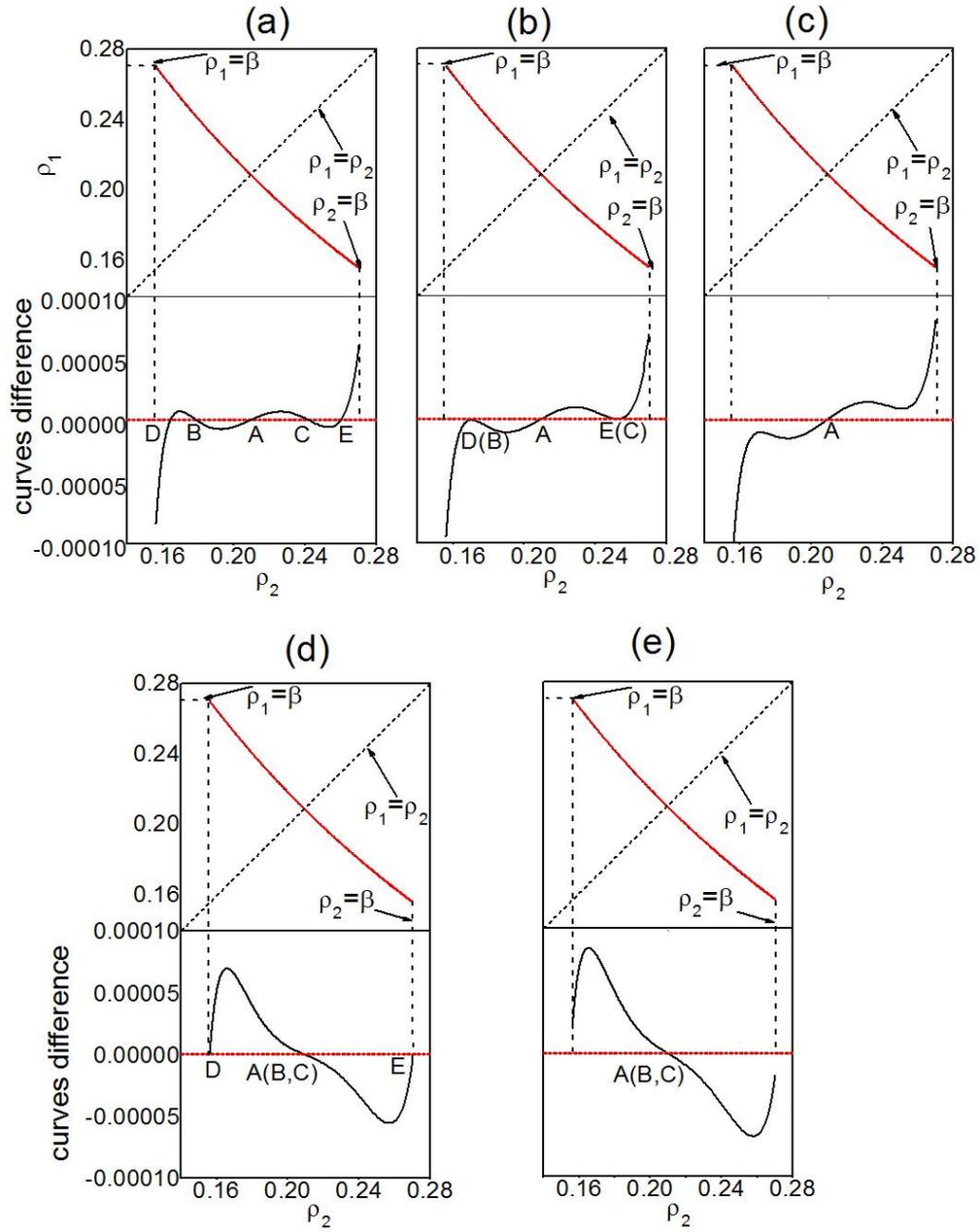

FIG. 5. The two curves of $\rho_1$ versus $\rho_2$ (top) and difference of the two curves (bottom). The parameter $\alpha = 0.8$, (a) $\beta = 0.27014$, (b) $\beta = 0.270123$, (c) $\beta = 0.2701$, (d) $\beta = 0.2702699$, (e) $\beta = 0.2703$.

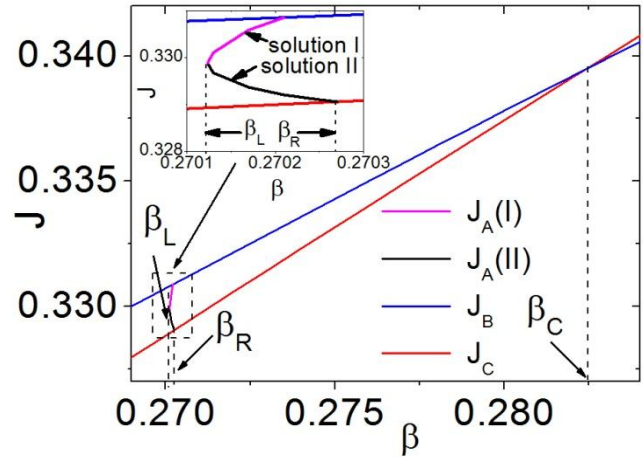

FIG. 6. The plot of $J_A$, $J_B$, and $J_C$ versus $\beta$, obtained from 5-cluster mean field analysis. The parameter $\alpha = 0.8$.

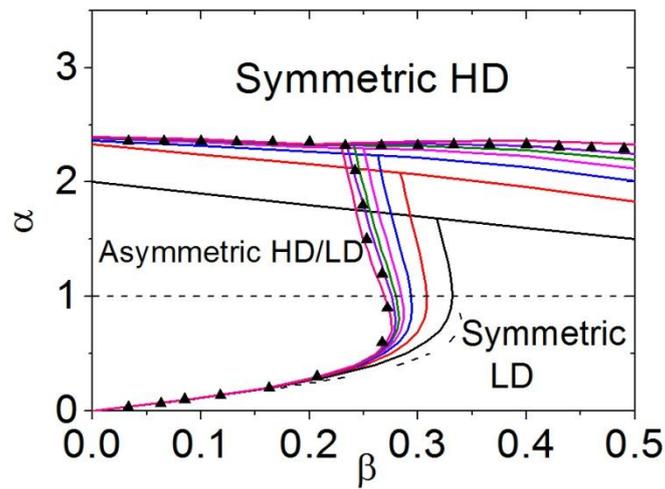

FIG. 7. Phase diagram of the system. Scattered data are from simulations (System size L=10000). Dash lines are from simple mean field analysis. Black line, red line, blue line, magenta line, olive line, violet line correspond to N-cluster mean field analysis with N=1-6, respectively. Pink lines are predicted results in the limit N→ ∞ based on exponential decay feature.

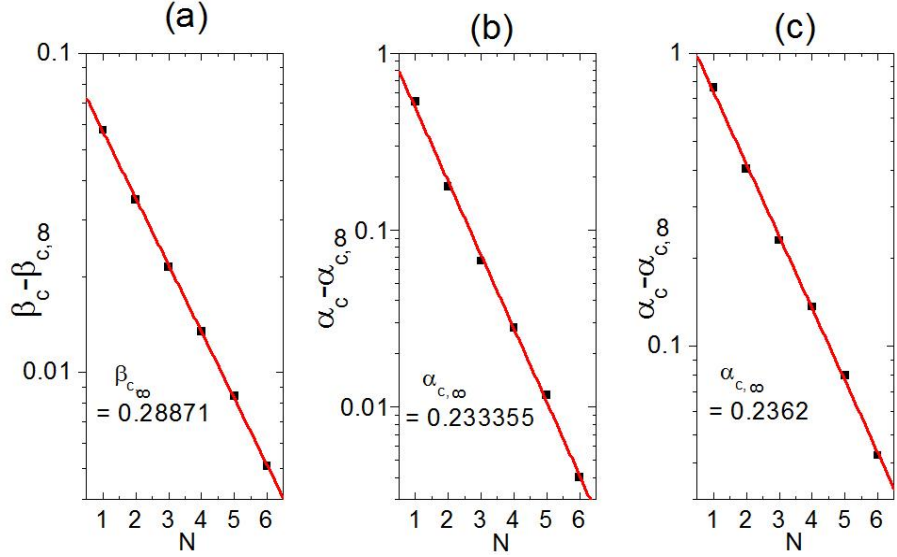

FIG. 8. Dependence of the boundary on N in N-cluster mean field analysis. (a) $\alpha = 0.9$, (b) $\beta = 0.2$, (c) $\beta = 0.4$. Scattered points are N-cluster mean field results, the red lines show fit of exponential decay.

## IV. Summary

To summarize, this paper studies symmetry breaking phenomenon in a bidirectional two-lane ASEP with narrow entrances, in which it is controversial whether asymmetric LD/LD phase exists or not. We point out that in previous simple mean field analysis, the necessary conditions have been wrongly regarded as the sufficient conditions. We demonstrate that the asymmetric LD/LD phase should not exist based on mean field analysis and current minimization principle. We observe an exponential decay feature in N-cluster mean field analysis, which has been used to predict phase boundary in the thermodynamic limit.

## Acknowledgements

This work is funded by the National Basic Research Program of China (No.2012CB725404) and the National Natural Science Foundation of China (Grant No. 11422221).